\documentclass[12pt]{revtex4}

\usepackage{psfig}
\usepackage{amsfonts}
\usepackage{amsbsy}
\usepackage{amssymb}

\def\staccrel#1#2{\mathrel{\mathop{#1}\limits_{#2}}}

\newcommand{\beq}{\begin{equation}}
\newcommand{\eeq}{\end{equation}}

\newcommand{\Ci}{{\cal{I}}}
\newcommand{\Cn}{{\cal{N}}}
\newcommand{\Cs}{\mbox{const.\,}}
\newcommand{\EA}{\textrm{E}}
\newcommand{\om}{\overline{m}}
\newcommand{\gb}{\bar{g}}
\newcommand{\smallspace}{\vspace{2.5ex}}

\newtheorem{theorem}{Theorem}

\begin{document}

\title{Probabilistic Regularization in Inverse Optical Imaging}

\author{Enrico De Micheli}
\affiliation{IBF -- Consiglio Nazionale delle Ricerche,
Via De Marini 6, 16149 Genova, Italy}

\author{Giovanni Alberto Viano}
\affiliation{Dipartimento di Fisica -- Universit\`a di Genova,
Istituto Nazionale di Fisica Nucleare, sez. di Genova, Via
Dodecaneso 33, 16146 Genova, Italy}

\begin{abstract}
The problem of object restoration in the case of spatially incoherent illumination is considered.
A regularized solution to the inverse problem is obtained through a probabilistic approach,
and a numerical algorithm based on the statistical analysis of the noisy data is presented.
Particular emphasis is placed on the question of the positivity constraint, which is incorporated
into the probabilistically regularized solution by means of a quadratic programming technique.
Numerical examples illustrating the main steps of the algorithm are also given.
\end{abstract}

\maketitle

\section{Introduction}
\label{introduction_section}

The inverse problem in optics consists of recovering the object by starting from
its image. It can be regarded as a backward channel communication problem: messages
can be conveyed back from the data set (the image) to reconstruct the signal (the object).
The length of these messages
is limited by the noise affecting the imaging process. This fact can be viewed as the necessity of truncating
the eigenfunction expansions associated with the Fredholm integral equation, which gives the mathematical
formulation of the inverse problem in optics. The latter can be formulated as follows \cite{Boccacci}:
Consider a one--dimensional object, illuminated by spatially incoherent radiation and imaged by a perfect
optical instrument (i.e., without focus error) with a rectangular aperture. If we use $f(x)$ to denote
the spatial radiance distribution in the object plane, then the noise free spatial radiance distribution
in the image plane is given by
\beq
\label{uno}
(Af)(y) = \int_{-1}^1 K(y-x)f(x)\,{\rm d}x = g(y)~,~~~~~-1\leqslant y \leqslant 1,
\eeq
where
\beq
\label{due}
K(x) = \frac{\sin^2(cx)}{\pi cx^2},~~~~~c = \frac{\pi}{R},
\eeq
$R$ being the Rayleigh resolution distance. Here we have assumed finite extent of the object and
the linear magnification of the optical instrument to be $+1$.

When $g(y)$ is given in the geometrical region of the image plane, i.e., the interval
$[-1,1]$, the problem of object restoration is equivalent to solving the
Fredholm integral equation of the first kind $Af=g$, where $A$ is a symmetric compact positive definite operator.
The solution of the integral equation is unique, i.e., the equation $Af=0$ has only the solution $f=0$.

A formal solution to the equation $Af=g$ can be given by means of an eigenfunction expansion, i.e.,
\beq
\label{tre}
f(x) = \sum_{k=1}^\infty \frac{g_k}{\lambda_k}\, \psi_k(x),
\eeq
where $g_k = (g,\psi_k)$ [$(\cdot,\cdot)$ denoting the scalar product in $L^2(-1,1)$], and $\psi_k(x)$ and $\lambda_k$
are, respectively, the eigenfunctions and the eigenvalues associated with the operator $A$. If we add
to the image $g$ a perturbation such as the one produced by the noise measurement, then
the equation $Af+n=\gb=g+n$, $n(y)$ being a function describing the noise or the experimental error,
generally has no solution. Accordingly, the expansion $\sum_{k=1}^\infty (\gb_k/\lambda_k)\psi_k(x)$,
[$\gb_k=(\gb,\psi_k)$] diverges if $\gb$ does not belong to the range of the operator $A$.
We are faced with the pathology of the ill--posedness of the problem, in the sense of Hadamard \cite{Hadamard}:
small perturbations of the data produce wide oscillations of the solutions. The problem requires regularization.
One of the most popular methods in use is due to Tikhonov \cite{Groetsch,Tikhonov} and consists of restricting
the solution space by imposing suitable global bounds on the solutions.
Within the framework of the eigenfunction expansions, Tikhonov's regularization provides a criterion
for a suitable truncation of the series. Furthermore, the function $f(x)$, which represents a spatial
radiance distribution, is a nonnegative function, and, consequently, a nonnegativity constraint must be
necessarily added in the mathematical formulation of the problem.

At this point it is worth noting that, although the $L^2$ space is the most natural ambient when the
method of the eigenfunction expansion is used and the Tikhonov's regularization is adopted,
the $L^1$ space could present several advantages, as we explain below.
First, the $L^1$ norm of the intensity is the energy radiated by the object, and therefore an equality of
the following type,
$\|f\|_{L^1(-1,1)}=\int_{-1}^1 |f(x)|\,{\rm d}x=E$, ($E$ is the energy radiated by the object), can clearly be interpreted
from a physical viewpoint. Analogously, the quantity $\int_{-1}^1 |Af-\gb|\,{\rm d}y$ represents the
statistical fluctuations of the energy of the image. Finally, the positivity constraint is sufficient to
restore the continuity of the inverse problem in the topology induced by the $L^1$ norm (see Appendix A).
Conversely, the positivity is not sufficient to regularize the problem in the $L^2$ topology.
All these considerations point towards choosing the $L^1$ norm in the mathematical formulation of the problem.
But it can easily be shown, by exhibiting explicit examples, that the continuity restored in the $L^1$ topology
is far from implying a continuity in the topology induced by the sup norm (see Appendix A).
Moreover, it should be observed that the positivity constraint is sufficient to restore the continuity in
the $L^1$ norm but not in the $L^2$ norm, simply because the $L^2$ convergence is more robust [note that
$f(x)$ has compact support] and represents precisely the minimal requirement in the actual numerical
calculations.
In view of this last consideration we are then led to choose the $L^2$ norm in the mathematical
formulation of the problem, and, consequently, the eigenfunction expansions can be used. However, we shall
not regularize the problem by using Tikhonov's variational method, which requires a bound of the
type: $\|f\|_{L^2(-1,1)}\leqslant M$, ($M$ is constant), whose interpretation is not clear from
the physical viewpoint.  Instead, we work out the problem with a probabilistic approach, which
makes it possible to split the Fourier coefficients of the noisy data function [i.e., $\gb_k = (\gb,\psi_k)$]
into two classes: one comprises those coefficients from which a significant amount of information
can be extracted; the other, those Fourier coefficients that can be regarded as random numbers
because the noise prevails on the information content. We can thus construct an approximation that satisfies
the requirements of the probabilistic regularization as explained in Subsection \ref{probabilistic_subsection}.
Questions related to the actual numerical computation of the probabilistically regularized solution
are discussed in Subsection \ref{correlation_section}, where an algorithm based on the analysis of the
autocorrelation function of the noisy data is presented.
However, at this stage, the positivity constraint still remains to be satisfied. This point is addressed
in Section \ref{numerical_section}, where the numerical issues concerning the construction
of a nonnegative approximation are discussed, and examples of numerical tests are also given.
Finally, in Appendix A, the ill--posedness of the problem in various topologies and the role played by
the positivity constraint will be illustrated.

\section{Probabilistic regularization and statistical methods}
\label{probabilistic_section}

\subsection{Probabilistic Regularization}
\label{probabilistic_subsection}

As remarked in Section \ref{introduction_section}, the ill--posed character of the inverse problem is derived precisely
from the fact that the data function $g(y)$ is corrupted by the noise that, hereafter, will be represented
by a bounded function $n(y)$, which is supposed to be integrable in the interval $[-1,1]$;
i.e., $\gb = g + n$. Furthermore, we set
\beq
\label{zero}
\left\|\gb-g\right\|_{L^2(-1,1)}=\left\|n\right\|_{L^2(-1,1)} \leqslant\epsilon.
\eeq
In general, the perturbation produced by the noise is such that the noisy data function $\gb$
does not belong to the range of the integral operator $A$, and, consequently, the eigenfunction expansion
$\sum_{k=1}^\infty(\gb_k/\lambda_k)\psi_k$ generally diverges.
Further, even assuming that the noise perturbation is gentle enough for $\gb$ to belong
to the range of the operator $A$, in this case the noise $n$ still precludes us from exactly knowing
the noiseless data function $g$. If we have two distinct data functions $\gb^{(1)}$ and $\gb^{(2)}$,
we may not attribute to them two distinct solutions $f^{(1)}$ and $f^{(2)}$, if the data distance is less than
$\epsilon$, i.e., if $\|\gb^{(1)}-\gb^{(2)}\|_{L^2(-1,1)} \leqslant\epsilon$. Accordingly, we can consider to be
distinguishable only those data such that $\|\gb^{(1)}-\gb^{(2)}\|_{L^2(-1,1)} > \epsilon$.
In conclusion, even if the series $\sum_{k=1}^\infty (\gb_k/\lambda_k)\psi_k$ converges in the $L^2$ norm, it is
meaningless to push the eigenfunction expansion beyond a certain value $k_0$ (i.e., a certain truncation point)
that depends on $\epsilon$.

The most intuitive and simple truncation method yielding a regularized solution
consists of writing an approximation of the following type,
\beq
\label{undici}
f_0(x) = \sum_{k=1}^{k_0(\epsilon)} \frac{\gb_k}{\lambda_k}\psi_k(x),
\eeq
where $k_0(\epsilon)$ is the largest integer such that $\lambda_k\geqslant\epsilon$
[assuming, in this particular case, the \textit{a priori} bound that the set of the input signals
belongs to the unit ball in $L^2(-1,1)$]. In fact,
it can be proved \cite{DeMicheli,Miller1,Miller2} that $f_0(x)$ converges weakly to $f(x)$ even if $\gb$ does not
belong to the range of the operator $A$, i.e.,
\beq
\label{dodici}
\lim_{\epsilon\rightarrow 0} \left(f-f_0,v\right)_{L^2(-1,1)}=0~,
~~~~[\forall v\in L^2(-1,1),~\|v\|_{L^2(-1,1)}\leqslant 1].
\eeq

\smallspace

\noindent
{\bf Remark}: If $\gb\in\textrm{range}\,(A)$, then it can be proved
by use of the Kolmogorov $\epsilon$--entropy theory \cite{Kolmogorov}
that $k_0(\epsilon)$ is strictly related to the maximal
length of the messages $L_{\max}(\epsilon)$ that can be conveyed back through the communication channel
associated to the operator $A$; in fact, we can prove that, for sufficiently small $\epsilon$ (see also,
Ref. \onlinecite{Scalas}),
\beq
\label{dieci}
L_{\max}(\epsilon) \simeq 2^{k_0(\epsilon)\log_2(1/\epsilon)}.
\eeq

\smallspace

But the approximation $f_0(x)$ presents several defects:
\begin{enumerate}
\item The solutions must be restricted to a bounded subset such as the unit ball in $L^2(-1,1)$ or,
equivalently, to a bounded subset of the type:
$\left\|f\right\|_{L^2(-1,1)} \leqslant M$, $M =$ constant, whose physical interpretation is not transparent,
as noted in Section \ref{introduction_section}.
\item Only a weak convergence of $f_0(x)$ to $f(x)$ is guaranteed, whereas at least the $L^2$--norm convergence
should be required for practical applications.
In addition, $f_0(x)$ does not generally satisfy the positivity constraint.
\item The truncation criterion $\lambda_k\geqslant\epsilon$ (or $\lambda_k\geqslant\epsilon/M$)
does not guarantee that the approximation $f_0(x)$ really does pick out the Fourier components of the noisy
data that are likely to carry exploitable information about the unknown solution and, at the same time,
reject the ones dominated by the noise.
\end{enumerate}

To overcome all these difficulties, we turn the problem in a probabilistic form.
With this in mind we rewrite integral equation (\ref{uno}) in the following form:
\beq
\label{pro_uno}
A\xi + \zeta =\eta,
\eeq
where $\xi$, $\zeta$ and $\eta$, which correspond to $f$, $n$ and $\gb$ respectively, are Gaussian weak random
variables in the Hilbert space $L^2 (-1,1)$ (see Ref. \onlinecite{Balakrishnan}).
A Gaussian weak random variable is uniquely
defined by its mean element and its covariance operator; in the present case we use
$R_{\xi \xi}$, $R_{\zeta \zeta}$, and $R_{\eta \eta}$ to denote the covariance operators of $\xi$, $\zeta$ and $\eta$,
respectively.

Next, we make the following assumptions:
\begin{itemize}
\item[I.] $\xi$ and $\zeta$ have zero mean; i.e. $m_\xi = m_\zeta = 0$.
\item[II.] $\xi$ and $\zeta$ are uncorrelated, i.e. $R_{\xi \zeta}$ = 0.
\item[III.] $R_{\zeta \zeta}^{-1}$ exists.
\end{itemize}
\noindent
The third assumption is the mathematical formulation of the fact that all the components of the data function
are affected by noise. As shown by Franklin [see formula (3.11) of Ref. \onlinecite{Franklin}],
if both signal and noise
satisfy assumptions (I) and (II), then $R_{\eta \eta}= A R_{\xi \xi} A^\star + R_{\zeta \zeta}$,
where $A^\star$ indicates the adjoint of $A$, and the
cross-covariance operator is given by: $R_{\xi \eta} = R_{\xi \xi}A^\star$. We also assume that $R_{\zeta \zeta}$
depends on a parameter $\epsilon$ that tends to zero when the noise vanishes, i.e., $R_{\zeta \zeta} = \epsilon^{2} N$,
where $N$ is a given operator (e.g., $N=I$ for white noise).

At this point,  we turn Eq. (\ref{pro_uno}) into an infinite sequence of one-dimensional equations by means of orthogonal
projections
\beq
\label{pro_cinque}
\lambda_k \xi_k+\zeta_k=\eta_k,~~~(k=1,2,...),
\eeq
where $\xi_k = (\xi, \psi_k)$, $\zeta_k = (\zeta, \psi_k)$, $\eta_k = (\eta,\psi_k)$ are Gaussian random variables.
Next, we introduce the variances $\rho_k^2 = (R_{\xi \xi} \psi_k, \psi_k)$,
$\epsilon^2 \nu_k^2 = (R_{\zeta \zeta} \psi_k, \psi_k)$,
$\lambda_k^2 \rho_k^2 + \epsilon^2 \nu_k^2 = (R_{\eta \eta} \psi_k, \psi_k)$,
without assuming that the Fourier components $\xi_k$ of $\xi$ (and analogously also for $\zeta_k$ and $\eta_k$)
are mutually uncorrelated. In view of assumptions (I) and (III) the following probability densities for $\xi_k$
and $\zeta_k$ can be assumed:
\beq
\label{pro_sei}
p_{\xi_k}(x)=\frac{1}{\sqrt{2\pi}\,\rho_k}\exp\left[-\left(\frac{x^2}{2\rho_k^2}\right)\right],~~~(k=1,2,\ldots),
\eeq
\beq
\label{pro_sette}
p_{\zeta_k} (x) = \frac{1}{\sqrt{2\pi}\,\epsilon\nu_k}\exp\left[ -\left (\frac{x^2}{2\epsilon^2\nu_k^2}\right)
\right],~~~(k=1,2,\ldots).
\eeq
Using Eq. (\ref{pro_cinque}), we can also introduce the conditional probability density
$p_{\eta_k} (y|x)$ of the random variable $\eta_k$ for fixed $\xi_k=x$, which reads as
\beq
\label{pro_otto}
p_{\eta_k} (y|x) = \frac{1}{\sqrt{2\pi}\,\epsilon\nu_k}\exp\left[-\frac{(y-\lambda_k x)^2}{2\epsilon^2\nu_k^2}
\right] = \frac{1}{\sqrt{2\pi}\,\epsilon\nu_k}
\exp\left[-\frac{\lambda_k^2}{2\epsilon^2\nu_k^2}\left (x-\frac{y}{\lambda_k}\right)^2\right].
\eeq
Let us now apply the Bayes formula that provides the conditional probability density of $\xi_k$ given $\eta_k$
through the following expression:
\beq
\label{pro_nove}
p_{\xi_k} (x|y) = \frac{p_{\xi_k}(x) p_{\eta_k} (y|x)}{p_{\eta_k} (y)}.
\eeq
Thus, if a realization of the random variable $\eta_k$ is given by $\gb_k$, formula (\ref{pro_nove}) becomes
\beq
\label{pro_dieci}
p_{\xi_k} (x|\gb_k) = A_k \exp\left[-\frac{x^2}{2\rho_k^2}\right]
\exp\left[-\frac{\lambda_k^2}{2\epsilon^2\nu_k^2}\left (x-\frac{\gb_k}{\lambda_k}\right)^2\right].
\eeq
Now, the amount of information on the variable $\xi_k$, which is contained in the variable $\eta_k$, can be evaluated.
We have \cite{Gelfand}
\beq
\label{pro_undici}
J(\xi_k, \eta_k) = - \frac{1}{2} \ln (1 - r_k^2),
\eeq
where
\beq
\label{pro_dodici}
r_k^2 = \frac{|\EA\left\{\xi_k\eta_k\right\} |^2}{\EA\left\{|\xi_k|^2\right\}\EA\left\{|\eta_k|^2\right\}} =
\frac{(\lambda_k\rho_k)^2}{(\lambda_k\rho_k)^2 + (\epsilon \nu_k )^2}.
\eeq
Thus, we obtain
\beq
\label{pro_tredici}
J (\xi_k,\eta_k) = \frac{1}{2} \ln\left (1 + \frac{\lambda_k^2 \rho_k^2}{\epsilon^2 \nu_k^2} \right ).
\eeq
Therefore, if $\lambda_k\rho_k < \epsilon \nu_k$, then $J(\xi_k, \eta_k) < (\ln 2)/2$.
Thus we are naturally led to introduce the following sets:
\beq
\label{pro_quattordicia}
{\cal I} = \left\{k\, :\, \lambda_k\rho_k \geqslant \epsilon\nu_k\right\},~~~~~
\Cn = \left\{k\, :\, \lambda_k\rho_k < \epsilon\nu_k\right\}.
\eeq
If we revert to the conditional probability density (\ref{pro_dieci}), this can be regarded as the product of two
Gaussian probability densities: $p_1 (x) = A_k^{(1)} \exp\left(-x^2/2\rho_k^2\right)$ and
$p_2 (x) = A_k^{(2)} \exp\left\{-(\lambda_k^2/2\epsilon^2\nu_k^2)\left[x-(\gb_k/\lambda_k)\right]^2\right\}$,
$(A_k=A_k^{(1)} \cdot A_k^{(2)})$, whose variances are given by $\rho_k^2$
and $(\epsilon\nu_k/\lambda_k)^2$, respectively.  Let us note that, if $k \in {\cal I}$, the variance associated with the density
$p_2 (x)$ is smaller than the corresponding variance of $p_1(x)$ and vice versa if $k \in \Cn$. Therefore
it is reasonable to consider as an acceptable approximation of $\langle\xi_k\rangle$ the mean value given by the density
$p_2(x)$ if $k \in {\cal I}$, but the mean value given by the density $p_1(x)$ if $k \in \Cn$. We can
then write the following approximation:
\beq
\label{pro_quindici}
\langle\xi_k\rangle = \left\{
\begin{array}{ll}
\gb_k/\lambda_k &~~~~~ k \in {\cal I} \\
0 &~~~~~ k \in \Cn
\end{array}
\right. .
\eeq
Consequently, given the value $\gb$ of the weak random variable $\eta$, we are led to the following estimate of $\xi$, which,
using the notation of Ref. \onlinecite{DeMicheli}, reads as
\beq
\label{novantasei}
\widehat{B} \gb = \sum_{k \in {\cal I}} \frac{\gb_k}{\lambda_k} \psi_k.
\eeq
Next, consider the global mean square error $\EA\left\{\|\xi - \widehat{B}\eta\|^2\right\}$ associated with the
operator $\widehat{B}$, introduced in (\ref{novantasei}). We can now prove the following theorem:
\begin{theorem}
\label{the:1}
If the covariance operator $R_{\xi\xi}$ is of trace class, and furthermore
$\lim_{k\rightarrow\infty}(\lambda_k\rho_k/\nu_k)=0$, then the following limit holds true:
\beq
\label{pro_diciassette}
\lim_{\epsilon\rightarrow 0}\, \EA\left\{\|\xi - \widehat{B}\eta\|^2\right\} = 0.
\eeq
\end{theorem}

\noindent {\bf Proof}: See Ref. \onlinecite{DeMicheli}.
\hfill$\square$

\subsection{Statistical Analysis of the Noisy Data}
\label{correlation_section}

The application of the results achieved in Subsection \ref{probabilistic_subsection} calls for statistical tools able to
determine the two sets $\Ci$ and $\Cn$. In this section this issue is discussed, and
the basic steps of a numerical algorithm for constructing the regularized solution $\widehat{B}\gb$ from
the noisy data $\gb$ are outlined. \\
Splitting the Fourier coefficients into the sets $\Ci$ and $\Cn$ can be performed by computing
the correlation function of the random variables $\eta_k$, i.e., the probabilistic counterpart of the
coefficients $\gb_k$,
\begin{equation}
\label{IV-1}
\Delta_{\eta}(k_1, k_2) =
\frac{\textrm{E}\{(\eta_{k_1}-\textrm{E}\{\eta_{k_1}\})(\eta_{k_2}-\textrm{E}\{\eta_{k_2}\})\}}
{\textrm{E}\{(\eta_{k_1}-\textrm{E}\{\eta_{k_1}\})^2\}^{1/2} \,\, \textrm{E}\{(\eta_{k_2}-\textrm{E}\{\eta_{k_2}\})^2\}^{1/2}}.
\end{equation}
In practice, only a finite realization $\{\gb_k\}_1^N$ of the random variables $\eta_k$ is usually available, from
which estimates $\delta_{\gb}$ of the autocorrelations can be obtained by regarding the data $\{\gb_k\}_1^N$
as a finite length record of a wide--sense--stationary random normal series \cite{Doob}.
In principle, the assumption of stationarity of the series $\{\eta_k\}$ is not strictly true, because the
moments of the random variables $\eta_k$ generally depend on $k$, but, from the practical point of view, this is
usually the only possible option. However, the stationarity assumption can be removed whenever estimates of
ensemble averages of the series $\{\eta_k\}$ can be computed.
Thus, by recalling that the $\eta_k$'s are normally distributed, by introducing the working hypothesis that
the process $\{\eta_k\}$ is stationary in wide sense \cite{Middleton},
i.e., $\Delta_{\eta}(k_1, k_2) = \Delta_{\eta}(k_1 - k_2)$, and by assuming that the ensemble contains no strictly
stationary subensembles that occur with probability other than zero or one, we can compute estimates of the
autocorrelation coefficients by means of the ergodic hypothesis \cite{Middleton}
equating ensemble and {\it time} (i.e., the index $k$ in our case) averages.

Among the numerous estimators of the autocorrelation function \cite{Jenkins}, one which is widely used
by statisticians is given by
\begin{equation}
\label{IV-2}
\delta_{\gb} (n) =
\frac{\strut\displaystyle
\sum_{k=1}^{N-n} (\gb_k - \langle\gb_k\rangle) (\gb_{k+n} - \langle\gb_{k+n}\rangle)}
{\left\{\strut\displaystyle
\sum_{k=1}^{N-n} (\gb_k - \langle\gb_k\rangle)^2 \sum_{k=1}^{N-n} (\gb_{k+n} - \langle\gb_{k+n}\rangle)^2\right\}^{1/2}},
~~~~n = 0, ..., N-1,
\end{equation}
where
\begin{equation}
\label{IV-3}
\langle\gb_k\rangle = \frac{1}{N-n} \sum_{k=1}^{N-n} \gb_k,~~~~~~
\langle\gb_{k+n}\rangle = \frac{1}{N-n} \sum_{k=1}^{N-n} \gb_{k+n}.
\end{equation}
\noindent
Equation (\ref{IV-2}), which is based on the scatter diagram of $\gb_{k+n}$ against $\gb_k$ for
$k = 1, .., N-n$, represents the maximum--likelihood estimate of the autocorrelation coefficients
of two random variables $\eta_k$ and $\eta_{k+n}$ whose joint probability distribution function
is bivariate normal.

To identify the structure of the series $\{\gb_k\}_1^N$ so as to separate the correlated components
from the the random ones, it is necessary to test whether $\delta_{\gb} (n)$ is effectively zero.
This question has been extensively discussed in Ref. \onlinecite{DeMicheli}.
Here, we briefly report on the main points. First we
assume that there exists an index $n_0$ such that for $n>n_0$, $\Delta_\eta(k_1-k_2)=\Delta_\eta(n)$ will
vanish. This index $n_0$ is actually recovered recursively as follows:
the series $\{\gb_k\}_1^N$ is first supposed to be purely
random, i.e., $n_0=0$, the standard error $\sigma_\delta (n;0)$ is computed, and the smallest index
$\overline{n}>0$ such that $|\delta_{\gb}(\overline{n})| > 1.96\,\sigma_\delta (n;0)$ is searched for.
If such an index $\overline{n}$ is found, it becomes the new candidate $n_0$, i.e., we set $n_0=\overline{n}$,
and the whole procedure is repeated until no new index $\overline{n}$ is found.
The {\em large-lag} standard error $\sigma_\delta (n;n_0)$ is evaluated by using the following formula, due
to Bartlett \cite{Bartlett}:
\beq
\label{IV-4}
\sigma_\delta (n,n_0) =  \left\{\frac{1}{N-n}
\left [1+2\sum_{v=1}^{n_0}\delta_{\gb}^2(v)\right]\right\}^{1/2}, ~~~~\mbox{for}~~ n > n_0.
\eeq
Formally, $n_0$ can be defined as
\begin{equation}
\label{IV-5}
n_0 = \max \, \{\overline{n} \geqslant 0 \, : \, \forall n \in (\overline{n},N-1], \,
\mid \delta_{\gb}(n)\mid < 1.96~ \sigma_\delta (n,\overline{n})\}.
\end{equation}
Accordingly, the set ${\bf Q}$ of the lags corresponding to autocorrelation values effectively different
from zero is defined as
\begin{equation}
\label{IV-6}
{\bf Q} = \{0 < n \leqslant n_0\, : \, \mid \delta_{\gb} (n) \mid > 1.96~\sigma_\delta(n,0)\}.
\end{equation}
Let $N_Q$ be the cardinality of $\bf Q$. From $\bf Q$ we can construct $N_Q$ families $F_i$ of pairs of
Fourier coefficients defined by
\begin{equation}
\label{IV-7}
F_i = \left\{(\gb_{k_i}, \gb_{k_i+n_i})\right\}_{k_i=1}^{(N-n_i)},~~~n_i \in {\bf Q},~~~i=1,\, ...,\, N_Q,
\end{equation}
from which the couples of coefficients $\gb_k$ that are likely to be correlated should be selected.
At this point, the Fourier coefficients that are correlated are determined in a unique way by means of the following
heuristic criterion: for any $n_i \in {\bf Q},\, i=1,...,N_Q$, we select the pair
$(\gb_{k^\star_i}, \gb_{k^\star _i+n_i})$ giving the maximum contribution to the corresponding autocorrelation
estimate $\delta_{\gb} (n_i)$, i.e., we define $k^\star_i$ as
\begin{equation}
\label{IV-9}
k^\star_i = \arg \max_{k\in [1,N-n_i]}\,\{|\gb_k\,\gb_{k+n_i}|\},~~~i=1,\, ...,N_Q.
\end{equation}
Accordingly, we can define the set of frequencies $\Ci$ that exhibit correlated Fourier coefficients
\begin{equation}
\label{IV-10}
\Ci = \{k^\star _i\}_1^{N_Q} \cup \{k^\star_i+n_i\}_1^{N_Q},
\end{equation}
where each element of $\Ci$ is counted only once. Finally, we can construct the approximation
\beq
\label{IV-11}
f_\Ci(x) = \sum_{k\in\Ci} \frac{\gb_k}{\lambda_k}\psi_k(x).
\eeq
In theory, that is, for $N\rightarrow\infty$, the $N_\Ci$ elements $k_i \in \Ci$ and the $N_Q$ numbers
$n_i \in {\bf Q}$ are mutually constrained. In fact, any two coefficients $k_\alpha, k_\beta \in \Ci$ must satisfy
the pairwise compatibility constraint requiring $|k_\alpha-k_\beta|\in {\bf Q}$. Moreover, it easy to see that
the number $N_\Ci$ of the admissible Fourier coefficients is combinatorially constrained by
$\frac{1}{2}\,\left[1+(1+8N_Q)^{1/2}\right] \leqslant N_\Ci \leqslant N_Q+1$.
In practice, that is, when the record length $N$ is finite, and in particular when the
signal-to-noise ratio (SNR) of the data $\gb$ is small, we cannot demand that all the compatibility constraints
be satisfied.
However, checking the number of compatibility constraints can provide us with a
confidence test on the reliability of the approximation $f_\Ci(x)$.

It is worth noting that, although we used the same notation, the set $\Ci$ of Eq. (\ref{IV-10})
can be different from the set $\Ci$ of Eq. (\ref{pro_quattordicia}). The former is actually the result
of an algorithm acting on a given set of data and can be thought of as a numerical realization of the
theoretical set $\Ci$ of (\ref{pro_quattordicia}). A similar role is played by the numerical approximation
$f_\Ci(x)$ with respect to the theoretical approximation $\widehat{B}\gb$ of Eq. (\ref{novantasei}).

\section{Numerical examples}
\label{numerical_section}

In Section \ref{probabilistic_section} we illustrated a statistical procedure that allows us to split
the Fourier coefficients into the two classes $\Ci$ and $\Cn$ and, accordingly, to write
the approximation $f_\Ci(x)$ [see formula (\ref{IV-11})], which, for the sake of convenience, can be rewritten as
\beq
\label{new2}
f_\Ci(x) = \sum_{m=1}^{m_\Ci(\epsilon)}\frac{\gb_m}{\lambda_m}\psi_m(x),
\eeq
where $m_\Ci(\epsilon)$ represents the cardinality of the set $\Ci$ and the eigenvalues $\{\lambda_k\}_{k\in\Ci}$
(and the corresponding eigenfunctions $\{\psi_k\}_{k\in\Ci}$) have been suitably relabelled
in a monotonic decreasing sequence.

In general, the approximation $f_\Ci(x)$ does not satisfy the positivity constraint.
How to incorporate effectively the positivity constraint into a regularizing scheme remains an open question,
which has been extensively discussed in the literature (see, for instance, Refs. \onlinecite{Bertero,Pike,McNally,Piana}
and the references therein).

\begin{figure}[ht]
\begin{center}
\leavevmode
\psfig{file=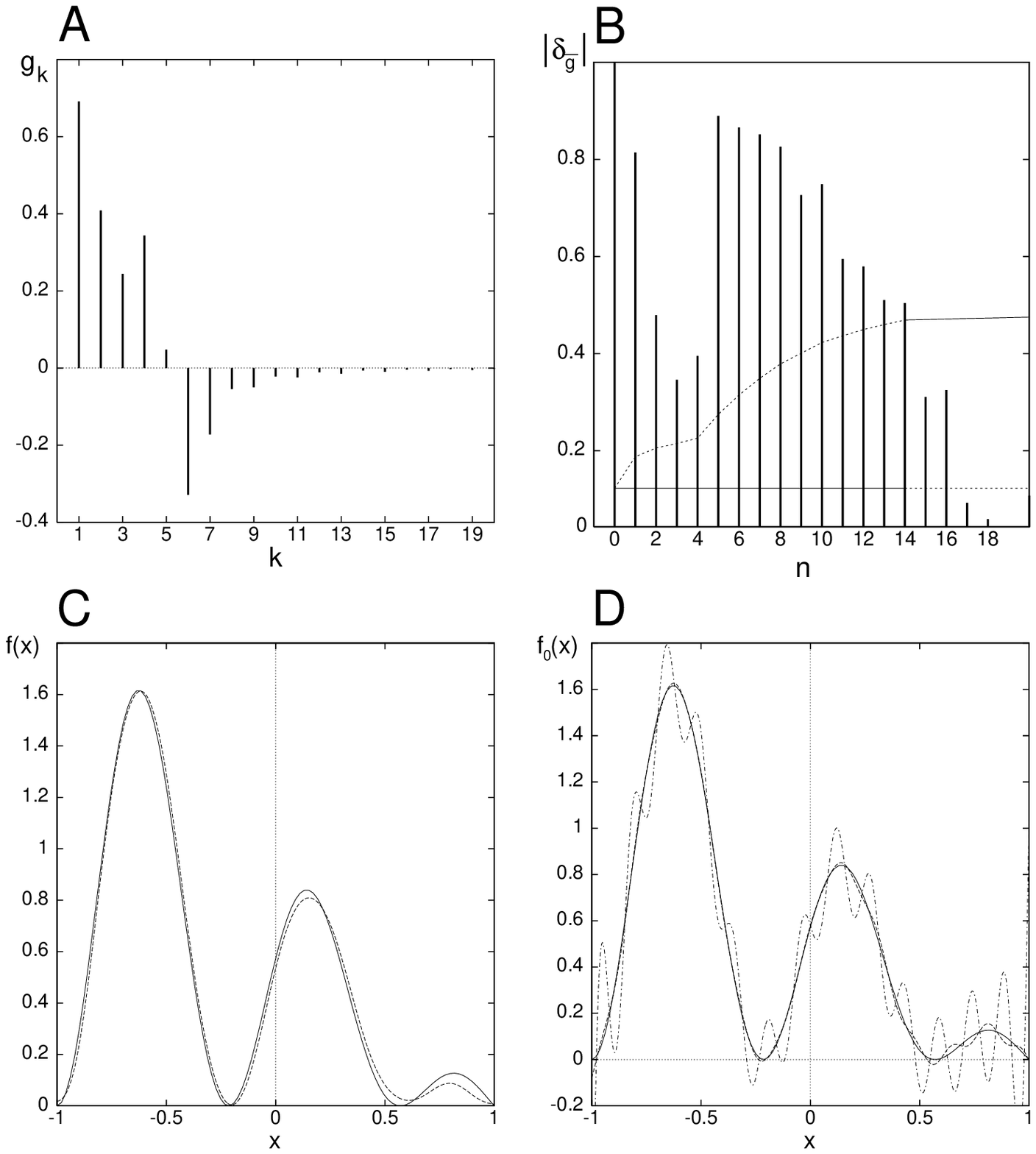,width=12cm}
\caption{\label{figura_1} \baselineskip=13pt Example 1:
$f_1(x)=(1-x)\sin^2[4(1+x)]$, $\epsilon=10^{-3}$, $c=20$. The
global SNR, defined as the ratio of the mean power of the
noiseless data to the noise variance, was SNR $\simeq
40\,\textrm{dB}$. A, Noiseless Fourier coefficients $g_k$. B,
Modulus of the autocorrelation function. From the analysis of
$\delta_{\bar{g}}(n)$ we have $n_0=14$, ${\bf
Q}=\{1,2,\ldots,13,14\}$. $\Ci=\{1,2,3,4,6,7,\ldots,14,15\}$, each
element with maximum inner compatibility, i.e., $13$. Horizontal
straight line, 95\% confidence limit $1.96\sigma_\delta(n;0)$ for
a purely random sequence. This limit was used to select the
elements of ${\bf Q}$ for $n\leqslant n_0 \equiv 14$ (solid part).
Curve, confidence limit $1.96\sigma_\delta(n;14)$ for $n>n_0$
(solid part) that we used for rejecting the autocorrelations that
are spuriously inflated by statistical fluctuations, whereas for
$n\leqslant n_0$ (dashed part) it shows only how the final
confidence limit $1.96\sigma_\delta(n;14)$ was reached during the
maximization procedure for setting $n_0$ [see text and, in
particular, Eq. (\ref{IV-5})]. C, Regularized solution. Solid
curve, actual solution $f_1(x)$. Dashed curve, reconstruction
$f_\Ci(x)$. D, Two examples of the approximation $f_0(x)$ [see Eq.
(\ref{undici})]. $k_0(\epsilon)$ was chosen as the largest integer
such that $\lambda_k\geqslant\epsilon/M$, where
$M=\|f_1(x)\|_{L^2(-1,1)}$, that is, in the current case, $k_0 =
29$. Solid curve, actual solution $f_1(x)$. Dashed curve, $f_0(x)$
computed with $k_0 = 27$ (see in particular the rightmost peak);
dotted--dashed curve, $f_0(x)$ computed with $k_0 = 28$. The
approximation $f_0(x)$ computed with $k_0 = 29$, as prescribed by
the above truncation criterion, is not displayed since it is
extremely different from the real solution.}
\end{center}
\end{figure}

If $f_\Ci(x)$ is not already a nonnegative function (see Fig. \ref{figura_1}), we can
aim at constructing a new positive approximation starting from the function $f_\Ci(x)$ itself.
The eventual negative part of $f_\Ci(x)$ is mainly due to
two factors: (i) the perturbation due to the noise that affects the coefficients $\{\gb_m\}_{m\in\Ci}$;
(ii) the error due to the truncation of the series expansion.
In the absence of additional prior information neither of these sources of error can be removed,
but we can nevertheless look for another approximation that is nonnegative
and \textit{similar} to $f_\Ci(x)$, according to suitable criteria.
With this in mind, our task is now to seek a positive regularizing solution of the following type:
\beq
\label{num1}
f_\Ci^{(p)}(x) = f_\Ci(x) + \sum_{m=m_\Ci(\epsilon)+1}^{\om_\Ci}d_m \psi_m(x),
\eeq
where $\om_\Ci$ is an integer parameter determining the maximum number of eigenfunctions $\psi_m(x)$ that can be
used to achieve the positive solution $f_\Ci^{(p)}(x)$ and the coefficients $d_m$ represent the unknown
coefficients to be determined by requiring the function $f_\Ci^{(p)}(x)$ to be nonnegative.
The function $f_\Ci^{(p)}(x)$ in Eq. (\ref{num1}) differs from the approximation $f_\Ci(x)$, resulting from
the analysis of the autocorrelation function, in the corrective finite linear combination of eigenfunctions
$\psi_m(x)$, whose purpose is to approximate and somehow compensate the errors leading to the negative part
of $f_\Ci(x)$.

This kind of approach quite naturally leads us to formulate the problem of finding the coefficients
$d_m$ as a mathematical programming problem \cite{Bazaraa}, i.e., choosing values of a set of variables
subject to various kinds of constraints placed on them. In particular, in our case we adopt a
quadratic programming scheme that can be summarized as follows: Minimize
\begin{eqnarray}
\label{num2}
F({\bf d}) &=& \|f_\Ci^{(p)}(x) - f_\Ci(x)\|_{L^2(-1,1)}^2, \label{num2} \\
{\bf d} &=& (d_{m_\Ci+1}, d_{m_\Ci+2}, \ldots, d_{\om_\Ci}), \label{numnumx}
\end{eqnarray}
subject to the constraints
\beq
\label{num3}
f_\Ci^{(p)}(x_i) \geqslant 0~,~~~~~i = 1, 2, \ldots, N_p,
\eeq
where $\{x_i\}_{i=1}^{N_p}$ is a given set of points distributed on the interval $[-1,1]$, and
\beq
\label{num4}
|\lambda_m d_m - \gb_m | \leqslant \epsilon_m~,~~~~~m = m_\Ci+1, \ldots, \om_\Ci,
\eeq
where $\epsilon_m$ represents an upper bound on the $m^{th}$ Fourier component of the noise $n(y)$, computed
with respect to the basis $\psi_m$. In practice only a global bound $\epsilon$ on the noise is usually available,
whereas the bounds $\epsilon_m$ on the single Fourier components are unknown. This leads us to substitute, a bit arbitrarily,
$\epsilon$ for the $\epsilon_m$'s, thus allowing the coefficients $d_m$ to vary on a wider range. However,
it will be shown in the numerical examples that this question is not a major issue, since the resulting products
$\lambda_m d_m$ generally differ from the corresponding $\gb_m$ much less than $\epsilon$.

Objective function (\ref{num2}) is of a type that reflects our strategy: We look for a positive solution that is closest,
in the sense of the $L^2$ norm, to the approximation $f_\Ci(x)$, which is not necessarily a positive solution to our
problem. Constraints (\ref{num3}) simply express the explicit requirement of positivity on a selected set of points.
Regarding the constraints (\ref{num4}), they basically require the coefficients $d_m$ to be compatible with the Fourier
coefficients of the data $\gb_m$ within the noise level. Concerning the actual numerical implementation of the
algorithm, eigenvalues and eigenfunctions of operator $A$ [see Eq. (\ref{uno})] were computed numerically
for different values of the parameter $c$ by using the Gauss--Legendre quadrature \cite{Numerical} and
subsequently by
diagonalizing the discretized problem by means of standard routines \cite{Numerical}. The constrained
optimization procedure was built around the routine E04NCF from the Nag Library.
For every test function $f(x)$,
the corresponding data function $g(y)$ was computed with Eq. (\ref{uno}), and then noise was added. For simplicity
we used only data corrupted by white noise simulated by computer generated random numbers uniformly distributed in the
interval $[-\epsilon,\epsilon]$.  However, provided the assumption of independence between $\xi$ and $\zeta$
(see Subsection \ref{probabilistic_subsection}), more general cases involving \textit{colored} noise
could be treated by using suitable methods such as \textit{prewhitening} transformations \cite{Box},
whose discussion is beyond the scope of this paper. Finally, the performance of the algorithm is evaluated by
direct comparison between the reconstructed solution and the true solution $f(x)$.

In Fig. \ref{figura_1} the main steps of the analysis outlined in Subsection \ref{correlation_section}
are shown. The sample function is $f_1(x)=(1-x)\sin^2[4(1+x)]$ with noise boundary
$\epsilon = 10^{-3}$ and corresponding SNR $\simeq 40$ dB.
The first noiseless Fourier coefficients $g_k$ of the image function, computed by using the kernel
$K(x,y)$ with $c=20$, are plotted in Fig. \ref{figura_1}A (see the legend for numerical details).
Figure \ref{figura_1}B shows the behaviour of the autocorrelation function $\delta_{\bar{g}}(n)$ along with
the lines that indicate the statistical confidence limits we used to discriminate whether
the autocorrelation function is substantially null.
Note that the autocorrelations at $n=15, 16$ have been correctly rejected, in spite of their quite large value,
since they were abnormally inflated by the autocorrelations with $n<15$.
The approximation $f_\Ci$ (dashed curve), computed from the set $\Ci$ obtained by analysis of the autocorrelation
shown in Fig. \ref{figura_1}B, and the actual object function $f_1(x)$ (solid curve) are displayed
in Fig. \ref{figura_1}C.
The excellent approximation supplied by $f_\Ci(x)$ is clearly evident. Moreover, the approximation does not
require any additional processing, since it is also nonnegative.

For the sake of comparison, the approximation $f_0(x)$, defined by Eq. (\ref{undici}),
is reported in Fig. \ref{figura_1}D. The criterion $\lambda_k\geqslant\epsilon/M$ with $M=\|f_1(x)\|_{L^2(-1,1)}$
gives $k_0(\epsilon)=29$ as truncation index. However, $f_0(x)$ computed with this value of $k_0$ (not plotted)
yields a very unsatisfactory approximation of the actual solution, and this failure can be ascribed
to the convergence, only of weak type, of the approximation $f_0(x)$ [see Eq. (\ref{dodici})].
In this case, the sole constraint on the norm of the solution is not sufficient for regularizing the
problem; therefore additional \textit{a priori} information, for instance, on the first derivative of the solution,
would be needed to achieve an acceptable approximation. Further prior information would lead to a smaller value
of $k_0(\epsilon)$ and, consequently, to good solutions, as shown in Fig. \ref{figura_1}D, where
an excellent reconstruction obtained with $k_0(\epsilon)=27$ (dashed curve) is shown. In contrast,
with $k_0(\epsilon)=28$ (dashed--dotted curve) wild oscillations start appearing.

\begin{figure}[ht]
\begin{center}
\leavevmode
\psfig{file=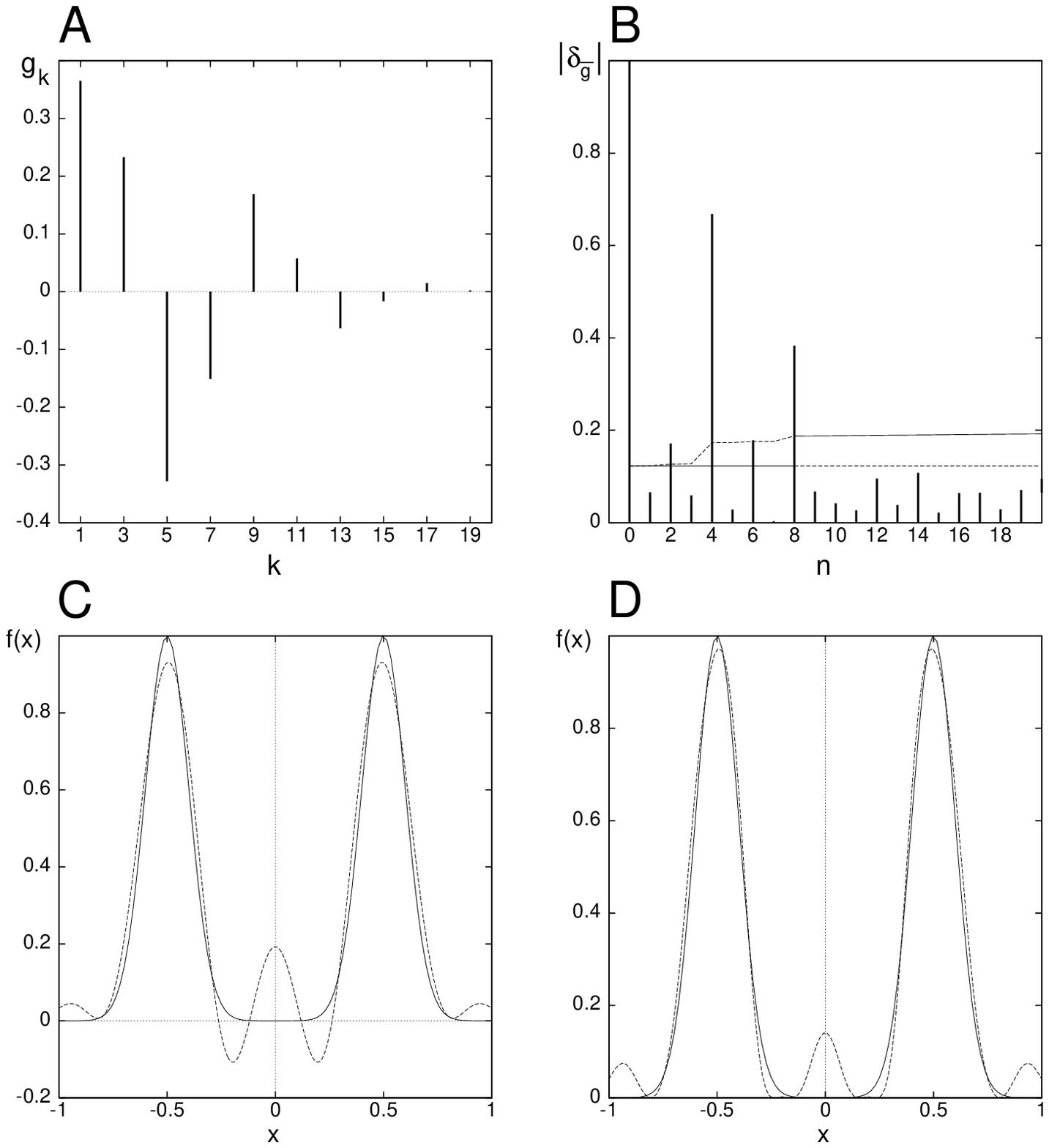,width=12cm}
\caption{\label{figura_2} \baselineskip=13pt Example 2:
$f_2(x)=\exp[-(x-x_0)^2/2\sigma^2]+\exp[-(x+x_0)^2/2\sigma^2]$
with $x_0=0.5$, $\sigma=0.1$, $\epsilon=3\times 10^{-2}$, SNR
$\simeq 6.2\, \textrm{dB}$, $c=20$. A, Noiseless Fourier
coefficients $g_k$. B, Modulus of the autocorrelation function.
$n_0 = 8$, ${\bf Q}=\{2,4,6,8\}$, $\Ci=\{1,3,5,7,9\}$. C,
Unconstrained regularized solution. Solid curve, actual solution
$f_2(x)$. Dashed curve, reconstruction $f_\Ci(x)$. D, Comparison
between the actual solution $f_2(x)$ and the constrained
regularized solution $f_\Ci^{(p)}(x)$ [see Eq. (\ref{num1})]. The
number of corrective eigenfunctions used was $\om_\Ci-m_\Ci=8$.
The positivity constraint was set over $N_p=64$ points of the
interval $[-1,1]$ [see relation (\ref{num3})]. After the
optimization procedure the coefficients $d_k$ were such that
$\max_{m_\Ci\leqslant m \leqslant \om_\Ci}|d_m
-\lambda_m\bar{g}_m|\simeq 0.01 < \epsilon$.}
\end{center}
\end{figure}

In general, $f_\Ci(x)$ does not satisfy the positivity constraint, in particular when the SNR becomes small,
and so the approximation $f_\Ci^{(p)}(x)$ must be computed.
The basic points of the algorithm for constructing the positive regularized solution $f_\Ci^{(p)}$ are summarized in
Fig. \ref{figura_2} for the sample function $f_2(x)=\exp[-(x-x_0)^2/2\sigma^2]+\exp[-(x+x_0)^2/2\sigma^2]$
with $x_0=0.5$ and $\sigma=0.1$. In this example, $\epsilon = 3\times 10^{-2}$, and SNR $\simeq 6.2$ dB.
Since the eigenvalues of the operator $A$ tend to decrease almost linearly with respect to the order
index when $k<4c/\pi$ (i.e., $k < 26$ in this example), whereas for $k>4c/\pi$ they go to zero exponentially fast
\cite{Gori}, we expect to recover quite accurately the object--function $f_2(x)$, which is
characterized as having the bulk of its information localized in the first values of $k$, even in the
presence of such a small SNR.
The analysis of the autocorrelation function $\delta_{\bar{g}}(n)$, shown in Fig. \ref{figura_2}B, leads
to selection of the Fourier components $\Ci=\{1,3,5,7,9\}$ with a high degree of confidence, since every element
of $\Ci$ satisfies all the compatibility constraints (see Subsection \ref{correlation_section}). Moreover, note
that only odd components were picked up, yielding an approximation with the same parity of the original
object function.
Although the reconstruction $f_\Ci$, illustrated in Fig. \ref{figura_2}C, is quite accurate, it presents
regions where it takes nonsense negative values. Figure \ref{figura_2}D shows the result of the
constrained optimization procedure we used to achieve a positive solution.
In this case only $\om_\Ci-m_\Ci=8$ corrective components have been used, while the positivity constraint has been
imposed on $N_p= 64$ equidistant points of the interval $[-1,1]$. It is worth noting that the parameter
$\om_\Ci$, which determines the number of corrective terms in the solution (\ref{num1}), is arbitrary and must
be set manually, with the only condition being that the constrained optimization problem has a nonempty feasible region.
Therefore different values of $\om_\Ci$ can lead to different positive problem solutions. In this example
we have shown the solution corresponding to the smallest value of $\om_\Ci$ which gives a minimum of the objective
function (\ref{num2}) compatible with constraints (\ref{num3}) and (\ref{num4}).

\begin{figure}[ht]
\begin{center}
\leavevmode \psfig{file=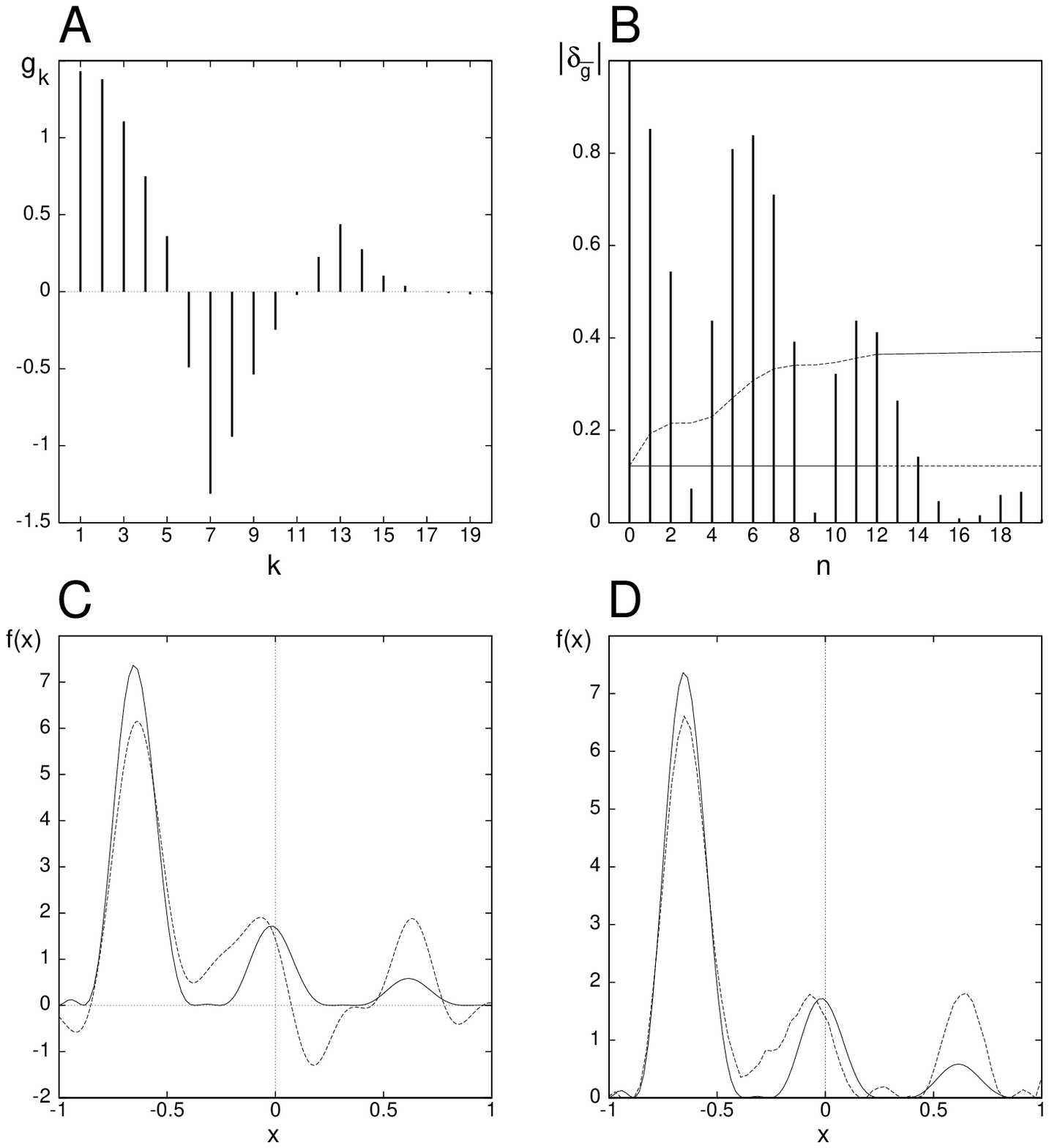,width=12cm}
\caption{\label{figura_3} \baselineskip=13pt Example 3: $f_3(x)=\sin^2[5(1-x)]\sin^2[5(1+x)][\exp(-3x)+\exp(-x)]$.
$\epsilon=5\times 10^{-2}$, SNR $\simeq 16.1 \textrm{dB}$, $c=20$.
A, Noiseless Fourier coefficients $g_k$.
B, Modulus of the autocorrelation function.
$n_0 = 12$, ${\bf Q}=\{1,2,4,5,6,7,8,10,11,12\}$,
$\Ci=\{1,2,3,7,8,9,13\}$. Each element of $\Ci$ has maximum inner compatibility, i.e., 6, with respect to the
set ${\bf Q}$.
C, Unconstrained regularized solution. Solid curve, actual solution $f_3(x)$.
Dashed curve, reconstruction $f_\Ci(x)$.
D, Comparison between the actual solution
$f_3(x)$ and the constrained regularized solution $f_\Ci^{(p)}(x)$ [see Eq. (\ref{num1})]. The number of
corrective eigenfunctions used was $\om_\Ci-m_\Ci=40$. The positivity constraint
was explicitly imposed on $N_p=64$ points of the interval $[-1,1]$ [see relation (\ref{num3})]. After the optimization
procedure the coefficients $d_k$ were such that
$\max_{m_\Ci\leqslant m \leqslant \om_\Ci}|d_m -\lambda_m\bar{g}_m|\simeq 0.02 < \epsilon$.}
\end{center}
\end{figure}

\begin{figure}[ht]
\begin{center}
\leavevmode \psfig{file=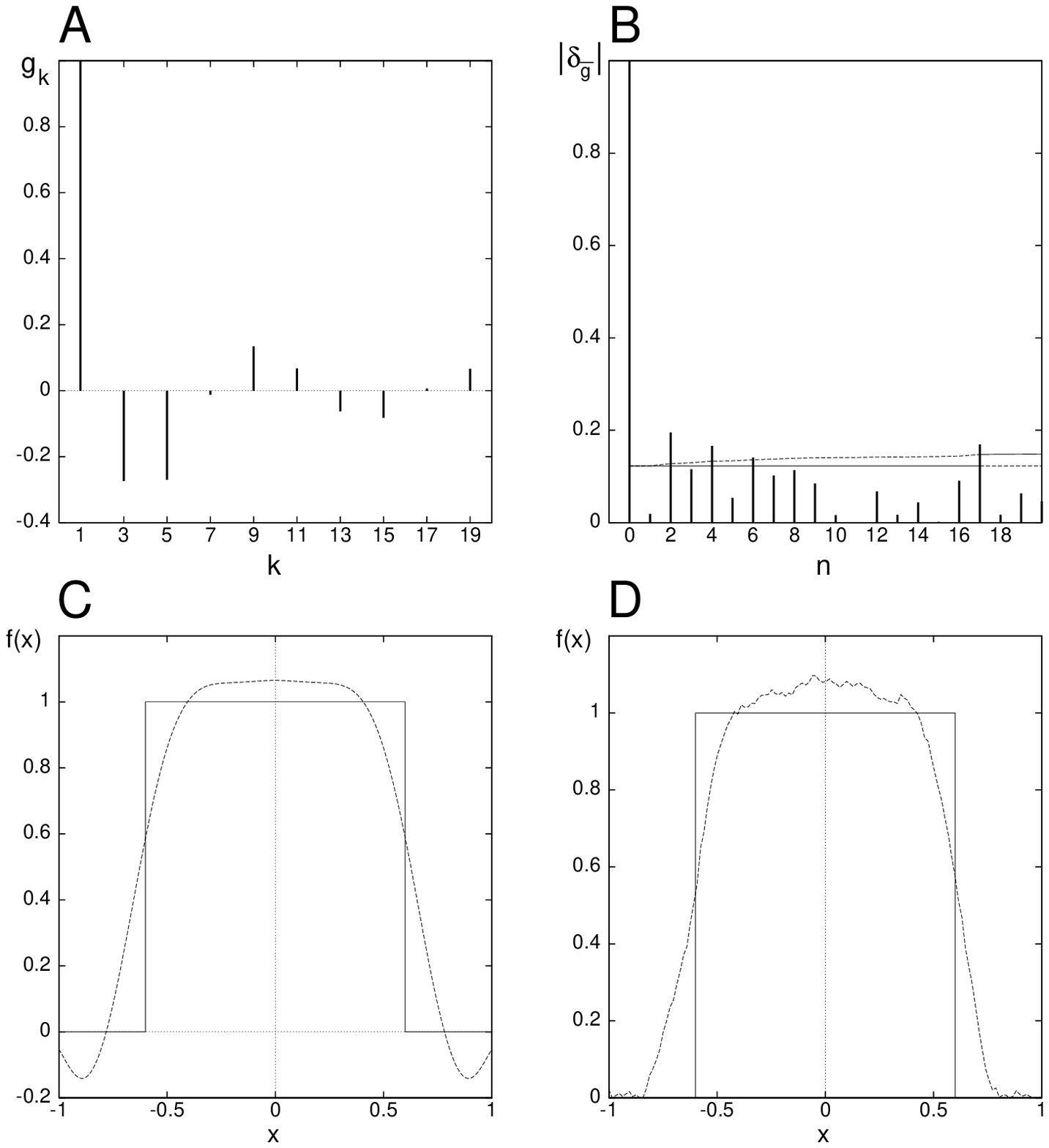,width=12cm}
\caption{\label{figura_4} \baselineskip=13pt Example 4: $f_4(x)=1$ for $-0.6\leqslant x\leqslant 0.6$ and null elsewhere.
$\epsilon=8\times 10^{-2}$, SNR $\simeq 3.2 \textrm{dB}$, $c=20$.
A, Noiseless Fourier coefficients $g_k$.
B, Modulus of the autocorrelation function.
$n_0=17$, ${\bf Q}=\{2,4,6,17\}$. The set ${\bf Q}$ led to selection of the
Fourier components $k=1,3,5,7,18$. However, the component at $k=18$ has the minimum compatibility index, i.e., 1,
which strongly indicates that the correlation at the lag $n=17$ was spuriously generated by the noise. Then,
$\Ci=\{1,3,5,7\}$.
C, Unconstrained regularized solution. Solid curve, actual solution $f_4(x)$.
Dashed curve, reconstruction $f_\Ci(x)$. D, Comparison between the actual solution
$f_4(x)$ and the constrained regularized solution $f_\Ci^{(p)}(x)$ [see Eq. (\ref{num1})]. The number of
corrective eigenfunctions used was $\om_\Ci-m_\Ci=70$. The positivity constraint
was set over $N_p=64$ points of the interval $[-1,1]$ [see relation (\ref{num3})]. After the optimization
procedure the coefficients $d_k$ were such that
$\max_{m_\Ci\leqslant m \leqslant \om_\Ci}|d_m -\lambda_m\bar{g}_m|\simeq 0.01 < \epsilon$.}
\end{center}
\end{figure}

Figures \ref{figura_3} and \ref{figura_4} sketch the results of two further examples.
In Fig. \ref{figura_3}C the unconstrained reconstruction $f_\Ci(x)$, which represents the starting
point of the optimization procedure, is less accurate than in the previous example.
By imposing the positivity constraint we can see that a higher accuracy in the reconstruction
of the true solution $f_2(x)$ was achieved in the regions where $f_\Ci(x)$ was already positive
(see the two leftmost peaks in Fig. \ref{figura_3}D).

Finally, Fig. \ref{figura_4} illustrates the restoration of an edge--type object.
Evidently, reconstructing discontinuous functions within a regularizing framework based on a truncated
Fourier expansion is a difficult task because of the Gibbs--like phenomenon. However,
the overall regularizing procedure,
including the positivity constraint on the solution, can provide an acceptable solution, as shown
in Fig. \ref{figura_4}D.
In this case the role of the parameter $\om_\Ci$ is quite critical.
The price to pay for reaching a reasonable solution is that many corrective terms must be used (in this example
$\om_\Ci = 70$) and also that different feasible values of $\om_\Ci$ can yield quite different reconstructions.

\newpage

\section*{Appendix A}
\label{appendix_section}

\setcounter{equation}{0}
\renewcommand{\theequation}{A.\arabic{equation}}

Let us consider the following set of functions $\{f_n(x)\}_{n=1}^\infty$:
\beq
\label{a1}
f_n(x) = \left(\frac{np+1}{2}\right)^{1/p}\left(1-|x|\right)^n~,~~~x\in [-1,1]~,~~~(p\geqslant 1).
\eeq
They are nonnegative functions if $x\in [-1,1]$ and we also have
\begin{eqnarray}
&&\sup_{x\in [-1,1]}|f_n(x)|=\left(\frac{np+1}{2}\right)^{1/p}\staccrel{\longrightarrow}{n\rightarrow\infty}\infty~,
~~~(p\geqslant 1), \\
&&\left\|f_n\right\|_{L^p} = 1~,~~~~~(p\geqslant 1), \\
\label{a4}
&&|Af_n| \leqslant \Cs \left(\frac{np+1}{2}\right)^{1/p}\frac{1}{n+1}.
\end{eqnarray}
From relation (\ref{a4}) it follows
$\lim_{n\rightarrow\infty}\left\|Af_n\right\|_{L^p} = 0$, if $p>1$.
Then, for any nonnegative function $f\in L^1(-1,1)$, we have
\beq
\label{a6}
\left\|Af\right\|_{L^1} \geqslant l \left\|f\right\|_{L^1}~,~~~~l = \inf_{x\in [-1,1]}\int_{-1}^1K(x,y)\,{\rm d}y > 0,
\eeq
and from relation (\ref{a6}) it follows that the operator $A^{-1}$ is continuous in the topology
of the $L^1$ norm.

We have thus proved that the positivity constraint restores continuity in the $L^1$ topology but
not in the $L^p$ topology if $p>1$.

\acknowledgments
We thank E.R. Pike for valuable discussions and helpful suggestions.

\end{document}